\documentclass[useAMS,fleqn, usenatbib, final]{mnras}
\usepackage{lipsum}
\usepackage{multirow}
\usepackage[inline]{enumitem}
\usepackage{import}
\usepackage{tabularx}
\usepackage{glossaries}
\usepackage[titletoc,title]{appendix}
\usepackage{graphicx}
\usepackage{amsmath}
\usepackage{amssymb}
\usepackage{amsfonts}
\usepackage{mathtools}
\usepackage{bm}
\usepackage[utf8]{inputenc}
\usepackage[table]{xcolor}
\usepackage{url}
\usepackage[normalem]{ulem}

\newcommand{\portsmouth}{Institute of Cosmology and Gravitation, University of Portsmouth, Portsmouth, PO1 3FX, UK}

\title[Signal plus noise inference with {\textsc{bilby}}]{A parametric signal plus noise inference framework for short duration non-Gaussian noise transients}

\author[Hoy]{
\parbox{\textwidth}{
C.~Hoy$^{1}$\thanks{charlie.hoy@port.ac.uk},
R.~Bondarescu$^{2,3}$,
A.~Lundgren$^{4,5}$,
L.K.~Nuttall$^{1}$
}
\vspace{0.2cm}\\
$^1$\portsmouth\\
$^2$Institut de Ci\`encies del Cosmos (ICCUB), Universitat de Barcelona (UB), c. Mart\'i i Franqu\`es, 1, 08028 Barcelona, Spain\\
$^3$Departament de F\'{\i}sica Qu\`antica i Astrof\'{\i}sica (FQA), Universitat de Barcelona (UB), 08028 Barcelona, Spain\\
$^4$Institut de Fisica d’Altes Energies, E-08193 Barcelona, Spain\\
$^5$Instituci\'o Catalana de Recerca i Estudis Avançats, E-08010 Barcelona, Spain
\vspace{-1.5em}
}
\date{Accepted XX. Received YY; in original form ZZ}
\pubyear{2026}

\begin{document}
\label{firstpage}
\pagerange{\pageref{firstpage}--\pageref{lastpage}}

\maketitle

\begin{abstract}
Gravitational waves are now routinely detected with ground-based observatories, and, through a process known as Bayesian inference, their source properties are inferred. However, terrestrial noise artifacts, often referred to as \emph{glitches}, commonly overlap astrophysical signals. This invalidates a fundamental assumption of gravitational wave analyses: the noise is no longer stationary and Gaussian. As a result, traditional techniques can provide biased inferences in realistic data. One method for mitigating the effect of glitches is to jointly analyse both the signal and noise in a single framework. In this work, we introduce {\textsc{bilby-antiglitch}} to infer the astrophysical signal properties in non-Gaussian noise. By additionally including a quasi-physical glitch model to describe short duration non-Gaussian noise transients, we show that unlike traditional techniques, we infer the true source properties of simulated signals contaminated with loud glitches. We also show that {\textsc{bilby-antiglitch}} prevents false violation claims of General Relativity, and validates the exceptional nature of gravitational wave signals in spurious data.
\vspace{1.5em}
\end{abstract}

\section{Introduction}

The Advanced LIGO~\citep{LIGOScientific:2014pky}, Virgo~\citep{VIRGO:2014yos} and KAGRA~\citep{KAGRA:2020tym} gravitational wave (GW) detectors have observed $\sim 400$ signals originating from compact binary coalescences~\citep[CBCs,][]{LIGOScientific:2026wfs}. The peak sensitivity of these detectors falls within the $20 - 1000\,\mathrm{Hz}$ frequency band, corresponding to the minimum of the noise power spectral density (PSD). Over short timescales, the detector noise is assumed to be Gaussian and stationary, allowing for the PSD to be estimated~\citep[see e.g.][]{Cornish:2014kda,Littenberg:2015kpb}. However, over longer durations~\citep[$\gtrsim 1$ minute,][]{Chatziioannou:2019zvs} the detector sensitivity is time dependent, meaning that the stationary assumption no longer holds~\citep{Mozzon:2020gwa,Edy:2021par}. In addition, non-Gaussian noise artifacts of terrestrial origin, hereafter referred to as \emph{glitches}, contaminate the GW strain data at a rate of $0.5-1.28$ per minute in each detector~\citep{LIGO:2024kkz,Ashton:2026seh}. When glitches overlap an astrophysical signal in both time and frequency, they can cause significant biases when inferring the source properties through Bayesian inference~\citep[see e.g.][]{Ghonge:2023ksb}. The clearest and most famous event that was significantly affected by a glitch was GW170817~\citep{LIGOScientific:2017vwq}. In the third GW observing run, 20\% of signals were polluted by glitches~\citep{Davis:2022ird}.

Glitches are classified by their distinct morphologies in time-frequency spectrograms. By leveraging these distinct shapes, machine learning frameworks like Gravity Spy~\citep{Zevin:2016qwy,Zevin:2023rmt} enable rapid, low-latency glitch classification. Examples of different glitch types include blips~\citep[narrow, tear-drop shapes lasting $\sim 5 - 10\,\mathrm{ms}$,][]{Cabero:2019orq}, tomtes~
citep[wider, low-frequency triangular profiles,][]{Glanzer:2022avx}, and koi fish~\citep[transients similar to blips but with excess noise on either side,][]{Bahaadini:2018git}. Since glitches are typically caused by local effects, they are statistically independent between detectors. This lack of correlation distinguishes glitches from true astrophysical signals, as only the latter is coincident in multiple detectors.

Glitch mitigation is vital for ensuring unbiased astrophysical inference. For example, \cite{Macas:2022afm} demonstrated that glitches can have a drastic effect on the sky localisation of the GW source where, in extreme circumstances, the true position can lie outside of the inferred 90\% credible interval. \cite{Lecoeuche:2026aix} also recently highlighted that short duration GW signals in particular are sensitive to blips, thunder, and fast-scattering, with parameter biases more severe when the glitch is near the merger time of the astrophysical signal. Several methods exist to subtract glitches from the data, including {\textsc{BayesWave}}, which models the glitch using a sum of sine-Gaussian wavelets~\citep{Cornish:2014kda,Littenberg:2015kpb,Cornish:2020dwh,Gupta:2024tve,Hourihane:2022doe}, and {\textsc{gwsubtract}}, which subtracts glitches based on data in auxiliary witness channels~\citep{Allen:1999wh,Davis:2018yrz}. Both these methods are preferred by the LIGO--Virgo-KAGRA collaboration~\citep[LVK,][]{LIGOScientific:2026wfs}, with {\textsc{BayesWave}} being consistently used since the second observing run~\citep{LIGOScientific:2017vwq,Pankow:2018qpo} and {\textsc{gwsubtract}} since the third observing run~\citep{KAGRA:2021vkt}.

To ensure that the underlying astrophysical signal is not removed along with the glitch, {\textsc{BayesWave}} performs a joint transdimensional inference of the signal and noise~\citep{Chatziioannou:2021ezd,Wijngaarden:2022sah}. However, it currently uses a simplified model for the astrophysical signal~\citep{Khan:2015jqa,Husa:2015iqa}, which neglects effects such as higher order multipole moments~\citep{Goldberg:1966uu,Thorne:1980ru} and spin-induced orbital precession~\citep{Apostolatos:1994mx}, and a data-driven noise model that is not derived from the phenomenology of real glitches. Therefore, it remains a potential concern that the sine-Gaussian noise model could remove astrophysical information from the GW signal, although this has been tested and shown to not be significant for several injections~\citep{Hourihane:2022doe}. Alternative methods for glitch mitigation exist in the literature, including jointly inferring the signal and noise with other glitch models, such as Gaussian processes~\citep{Ashton:2022ztk,Emma:2026zel}, machine learning algorithms~\citep{Malz:2025xdg}, scattering light and wavelet models~\citep{Udall:2025bts}, applying first order corrections to the likelihood to handle PSD variations~\citep{Zackay:2019kkv,Mozzon:2020gwa}, defining an alternative hyperbolic likelihood~\citep{Sasli:2026pds}, and using machine learning to subtract glitches based on auxiliary witness channels~\citep{Macas:2023wiw}.

In this work, we combine the core concept of jointly inferring the signal and noise in {\textsc{BayesWave}} with {\textsc{bilby}}; {\textsc{bilby}} has been used by the LVK to extract the source properties for all GW signals since the third observing run~\citep{KAGRA:2021vkt}. Unlike previous~\citep{Ashton:2022ztk,Malz:2025xdg,Udall:2025bts} and also future in-preparation implementations~\citep{Cheung:2026aaa}, we adopt a quasi-physical parameterised glitch model to describe short duration non-Gaussian noise transients, and interact with traditional off-the-shelf nested samplers. This removes the possibility of overfitting the data and provides improved interpretability with known failure modes. By implementing our methodology into {\textsc{bilby}}, our algorithm can trivially interface with the latest and most accurate waveform models and, thanks to its modular design, new glitch models can be added when available. To verify our framework, named {\textsc{bilby-antiglitch}}, we analyse a simulated signal and show that we recover the true source properties of the astrophysical signal injected into non-Gaussian data. We further highlight that, unlike traditional methods, {\textsc{bilby-antiglitch}} resolves possible false deviations of General Relativity that arise from spurious data. We finally apply {\textsc{bilby-antiglitch}} to GW250114\_082203~\citep{LIGOScientific:2025rid} and GW200129\_065458~\citep{KAGRA:2021vkt}. For GW250114\_082203, we obtain consistent results with the LVK and no significant evidence for glitches in the data. For GW200129\_065458 we show that the precession measurement reported in~\cite{Hannam:2021pit} is robust to possible short duration non-Gaussian noise transients.

In Section~\ref{sec:bayesian_inference} we discuss the methodology behind the joint signal plus noise inference in {\textsc{bilby-antiglitch}}. In Section~\ref{sec:verification} we verify the {\textsc{bilby-antiglitch}} algorithm with a simulated GW signal, and discuss the implications of using standard {\textsc{bilby}}. In Section~\ref{sec:application} we apply {\textsc{bilby-antiglitch}} to GW250114\_082203 and GW200129\_065458, and finally we conclude in Section~\ref{sec:discussion}.

\section{Bayesian inference} \label{sec:bayesian_inference}

Being able to extract the source properties from an observed GW signal enables key astrophysical insights into the nature of black holes~\citep[see e.g.][]{LIGOScientific:2026ctl} and their population properties~\citep[see e.g.][]{LIGOScientific:2025brd,LIGOScientific:2025rid}. Bayesian inference is generally used to estimate the source properties $\theta$ from some data $d$ given a model $m$. The parameters are represented by the \emph{posterior probability distribution}, and calculated through Bayes theorem,

\begin{equation}
    P(\theta | d, m) = \frac{\mathcal{L}(d | \theta, m)\, \Pi(\theta | m)}{\zeta}
\end{equation}
where $\mathcal{L}(d | \theta, m)$ is the probability of the data given the parameters and model, otherwise known as the likelihood, $\Pi(\theta | m)$ is the probability of the parameters given the model, otherwise known as the prior, and $\zeta = \int(\mathcal{L}(d | \theta, m)\, \Pi(\theta | m)\, d\theta$ is the probability of the data given the model, otherwise known as the evidence.

Assuming an astrophysical signal $h$ with true parameters $\hat{\theta}$, the GW data $d = h + n$ where $n$ is the instrumental noise. Assuming the noise is Gaussian and stationary, the likelihood takes the simplified Whittle form~\citep{Veitch:2014wba},

\begin{equation} \label{eq:likelihood}
    \mathcal{L}(d | \theta, m) = \exp\left(-\frac{1}{2}\langle d - m(\theta) | d - m(\theta) \rangle \right),
\end{equation}
where $\langle a | b \rangle$ denotes the inner product between two frequency series $a(f)$ and $b(f)$~\citep{Finn:1992wt},

\begin{equation}
    \langle a | b \rangle = 4\mathrm{Re}\int{df \frac{a(f)b^{*}(f)}{S_{n}(f)}},
\end{equation}
and $S_{n}(f)$ is the PSD of the noise.

Although the likelihood is well known, it is difficult to analytically compute the posterior distribution. As such, stochastic sampling algorithms, such as nested sampling~\citep{Skilling:2004pqw, Skilling:2006gxv} and Markov-Chain Monte-Carlo~\citep{metropolis1949monte}, are often used to draw samples from the unknown posterior distribution, although other approaches have also been proposed~\citep{Pankow:2015cra,Lange:2018pyp,Gabbard:2019rde,Green:2020dnx,Green:2020hst,Dax:2021tsq,Tiwari:2023mzf,Fairhurst:2023idl,Dax:2024mcn,Raymond:2024xzj,Roulet:2024hwz,Mushkin:2025yks,Williams:2025szm,Roulet:2026mzz}. Numerous software packages are available to perform Bayesian inference for GW astronomy~\citep{Ashton:2018jfp,Biwer:2018osg,Romero-Shaw:2020owr}, but in this work we use {\textsc{bilby}}~\citep{Ashton:2018jfp,Romero-Shaw:2020owr} and the {\textsc{dynesty}}~\citep{Speagle:2019ivv} nested sampler. Since {\textsc{bilby}} samples Eq.~\ref{eq:likelihood}, it currently assumes the noise is stationary and Gaussian, see e.g.~\cite{LIGOScientific:2026wfs}.

\subsection{The signal-only hypothesis}

A non-eccentric astrophysical signal from the merger of two black holes can be described by a fifteen dimensional model $m =\mathfrak{M}(\theta)$; two parameters describing the component masses $m_{1}$ and $m_{2}$, six spin parameters characterising the spin vectors of each black hole $\boldsymbol{\chi}_{1}$, $\boldsymbol{\chi}_{2}$, four parameters describing the sky location, luminosity distance, and the inclination angle of the source, and three describing the geocentric merger time, phase and polarization of the GW signal. Although formally described by fifteen parameters, mass-weighted effective parameters -- such as $\chi_{\mathrm{eff}}$~\citep{Ajith:2009bn} and $\chi_{\mathrm{p}}$~\citep{Schmidt:2014iyl} -- are more readily extracted from the data. $\chi_{\mathrm{eff}}$ describes the spin aligned with the orbital angular momentum, and it is defined as,

\begin{equation}
    \chi_{\mathrm{eff}} = \frac{\left(m_{1}\boldsymbol{\chi}_{1} + m_{2}\boldsymbol{\chi_{2}}\right)\cdot\mathbf{\hat{L}}}{m_{1} + m_{2}}.
\end{equation}
$\chi_{\mathrm{eff}}$ ranges between $-1$ and $1$ where $-1$ implies the spins are anti-aligned and $1$ means the spins are aligned with the orbital angular momentum.

Assuming the Whittle likelihood, Eq.~\ref{eq:likelihood} simplifies to,

\begin{equation}
    \mathcal{L}(d | \theta, m) = \exp\left(-\frac{1}{2}\langle n | n\rangle - \langle \delta \mathfrak{M} | n\rangle -\frac{1}{2} \langle \delta\mathfrak{M} | \delta\mathfrak{M} \rangle)\right),
\end{equation}
where $\delta \mathfrak{M} = h - \mathfrak{M}(\theta)$. If the model is a perfect description of the astrophysical signal, such that $\delta \mathfrak{M} = 0$ at $\theta = \hat{\theta}$, the peak likelihood simplifies to,

\begin{equation} \label{eq:gaussian_noise}
    \mathcal{L}(d | \theta, m) = \exp\left(-\frac{1}{2} \langle n | n \rangle \right).
\end{equation}
This demonstrates the fundamental assumption of the Whittle likelihood: the residual is assumed to follow a Gaussian distribution. In reality, a model is not a perfect description of the astrophysical signal, and any deviations from the peak likelihood is caused by random Gaussian noise fluctuations, $\langle \delta \mathfrak{M} | n \rangle \neq 0$, and model systematics, $\langle \delta\mathfrak{M} | \delta\mathfrak{M} \rangle \neq 0$. 

In GW astronomy, numerous models are available to describe binary black hole mergers. In the latest GW transient catalog~\citep{LIGOScientific:2026wfs}, the LVK used two models from the phenomenological family: {\texttt{IMRPhenomXPHM}}~\citep{Pratten:2020ceb,Colleoni:2024knd} and {\texttt{IMRPhenomXPNR}}~\citep{Hamilton:2025xru}, one from the effective-one-body family: {\texttt{SEOBNRv5PHM}}~\citep{Ramos-Buades:2023ehm}, and one from the NRSurrogate family: {\texttt{NRSur7dq4}}~\citep{Varma:2019csw}.

\subsection{The signal plus noise hypothesis}

When the noise is non-Gaussian, for example when a glitch $g$ overlaps the astrophysical signal in both time and frequency, the GW strain data is $d = h + n + g$. The Whittle likelihood is no longer valid as the true statistical distribution of the noise has heavier tails than the assumed Gaussian model. In other words, extreme outliers in the noise happen more often than the Gaussian assumption would predict. As such, we expect to observe significant biases when performing Bayesian inference under the signal-only hypothesis. For example, even if the model is a perfect description of the true underlying astrophysical signal, such that $\delta\mathfrak{M} = 0$, the Whittle likelihood in non-Gaussian noise simplifies to,

\begin{equation} \label{eq:non_gaussian_noise}
    \mathcal{L}(d | \theta, m) = \exp\left(-\frac{1}{2} \langle n | n \rangle - \langle n |g\rangle - \frac{1}{2} \langle g |g\rangle\right).
\end{equation}
Note that when $g = 0$, the fundamental Gaussian distribution in Eq.~\ref{eq:gaussian_noise} is recovered.

However, if we describe the astrophysical signal with the model $\mathfrak{M}$ and the non-Gaussian noise transient with the model $\mathcal{G}$, such that $m = \mathfrak{M} + \mathcal{G}$, we recover the fundamental assumption in the Whittle likelihood: any non-Gaussianities in the data can be removed alongside the astrophysical signal when subtracting the model $m$ from the data $d$. In other words, if $h = \mathfrak{M}(\theta)$ and $g = \mathcal{G}(\beta)$, Eq.~\ref{eq:gaussian_noise} is recovered. The downside is the increase in the dimensionality of the parameter space from $\theta$ for the signal-only hypothesis, to $\lambda = \{\theta, \beta\}$ where $\beta$ describes the parameters of the glitch model $\mathcal{G}$.

Several glitch models exist in the literature including the quasi-physical glitch model {\textsc{AntiGlitch}}~\citep{Bondarescu:2023jcx}. Although other studies have used data-driven glitch models, such as Gaussian Processes~\citep{Ashton:2022ztk} and continuous wavelets~\citep{Cornish:2014kda,Littenberg:2015kpb,Cornish:2020dwh,Gupta:2024tve,Hourihane:2022doe}, {\textsc{AntiGlitch}} adopts a parametric approach. This has several advantages, including reducing the possibility of overfitting the data and improving the glitch interpretability. In the frequency-domain, {\textsc{AntiGlitch}} is parameterised by five parameters per interferometer: two describing the peak frequency $f_{0}$ and the inverse bandwidth squared $\gamma$ of the glitch, and three describing the amplitude $A$, phase $\varphi$ and central time of the glitch $t_{0}$. The model is simply,

\begin{figure}
    \includegraphics[width=0.49\textwidth]{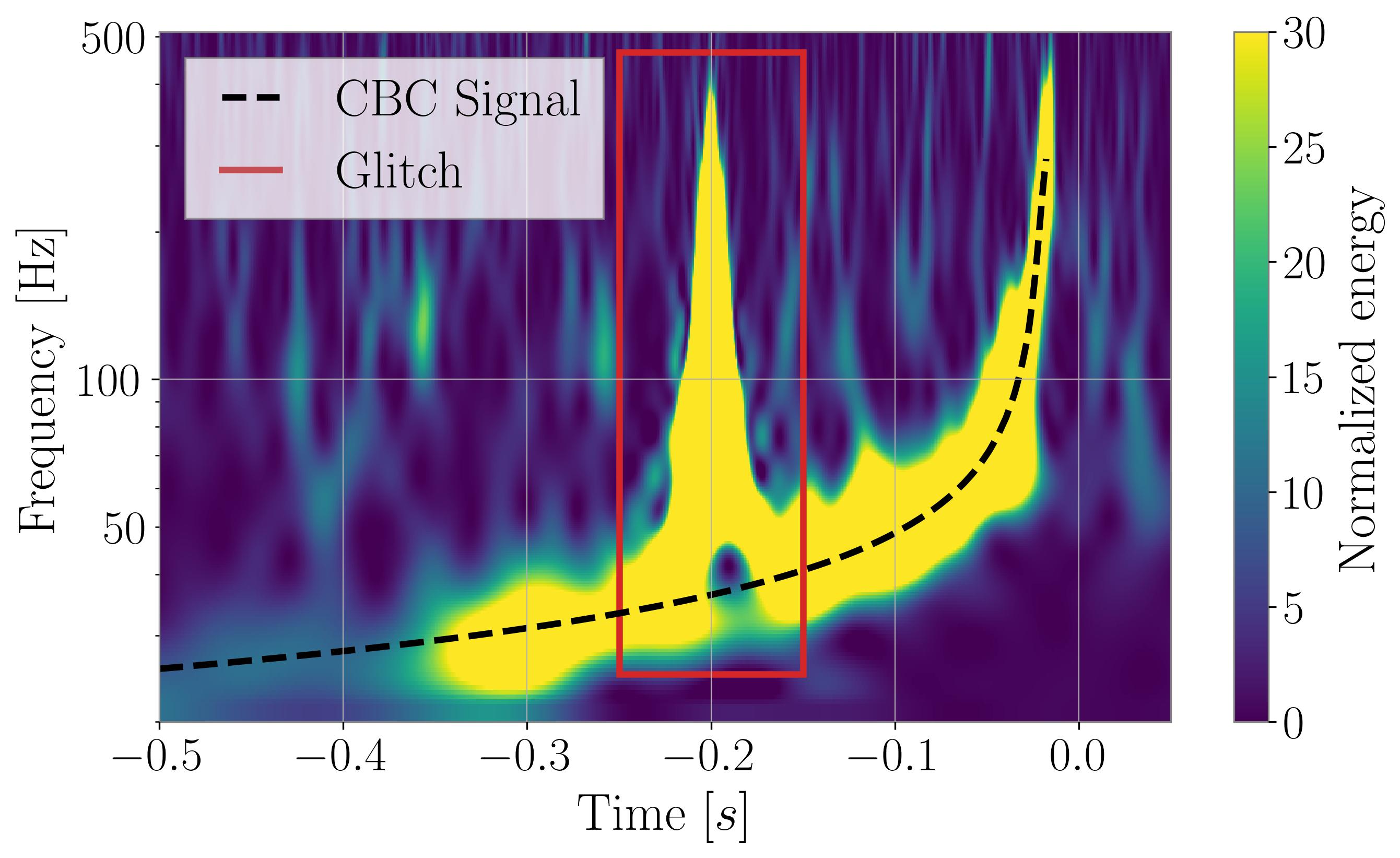}
    \caption{The spectrogram of a simulated GW signal in the presence of non-Gaussian noise, which was used to verify our signal plus noise inference pipeline. The dashed line shows the frequency track of the simulated GW150914-like signal, and the solid box isolates the blip glitch $0.2$ seconds before the merger.}
    \label{fig:verification_spectrogram}
\end{figure}

\begin{figure*}
    \includegraphics[width=0.95\textwidth,trim={0cm 9cm 0cm 0cm}, clip]{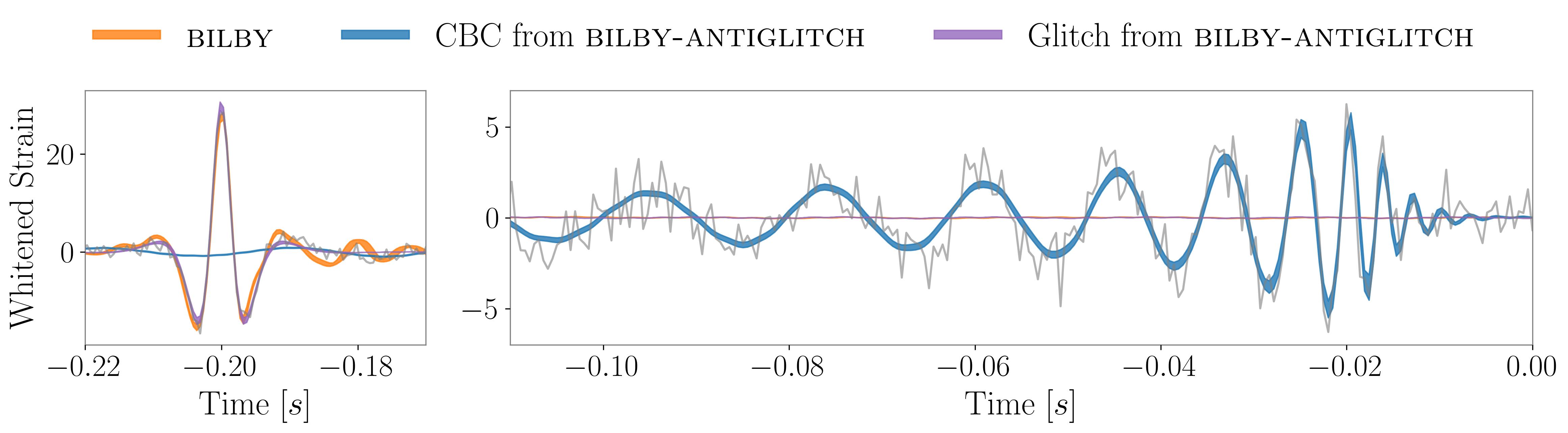}
    \includegraphics[width=0.48\textwidth]{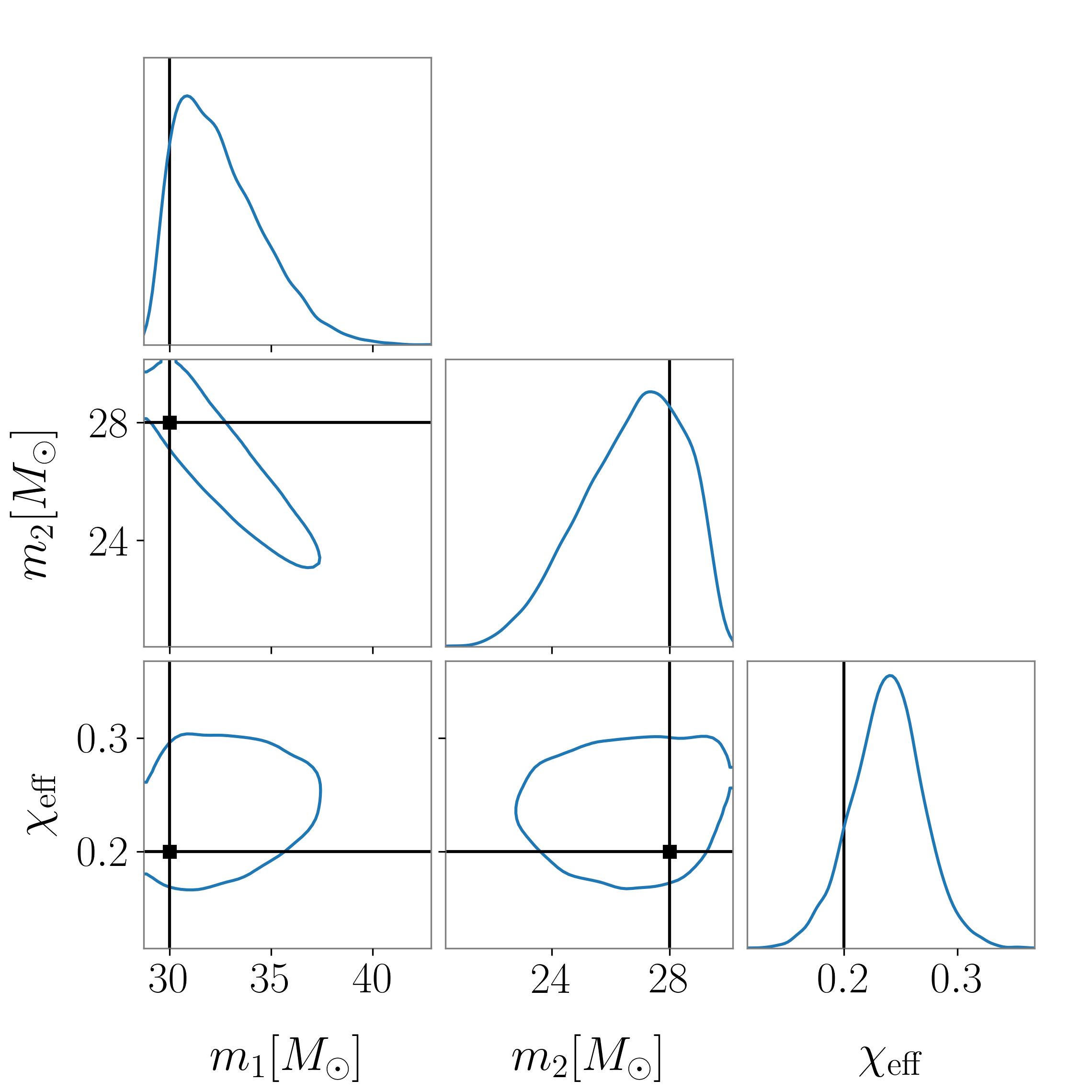}
    \includegraphics[width=0.48\textwidth]{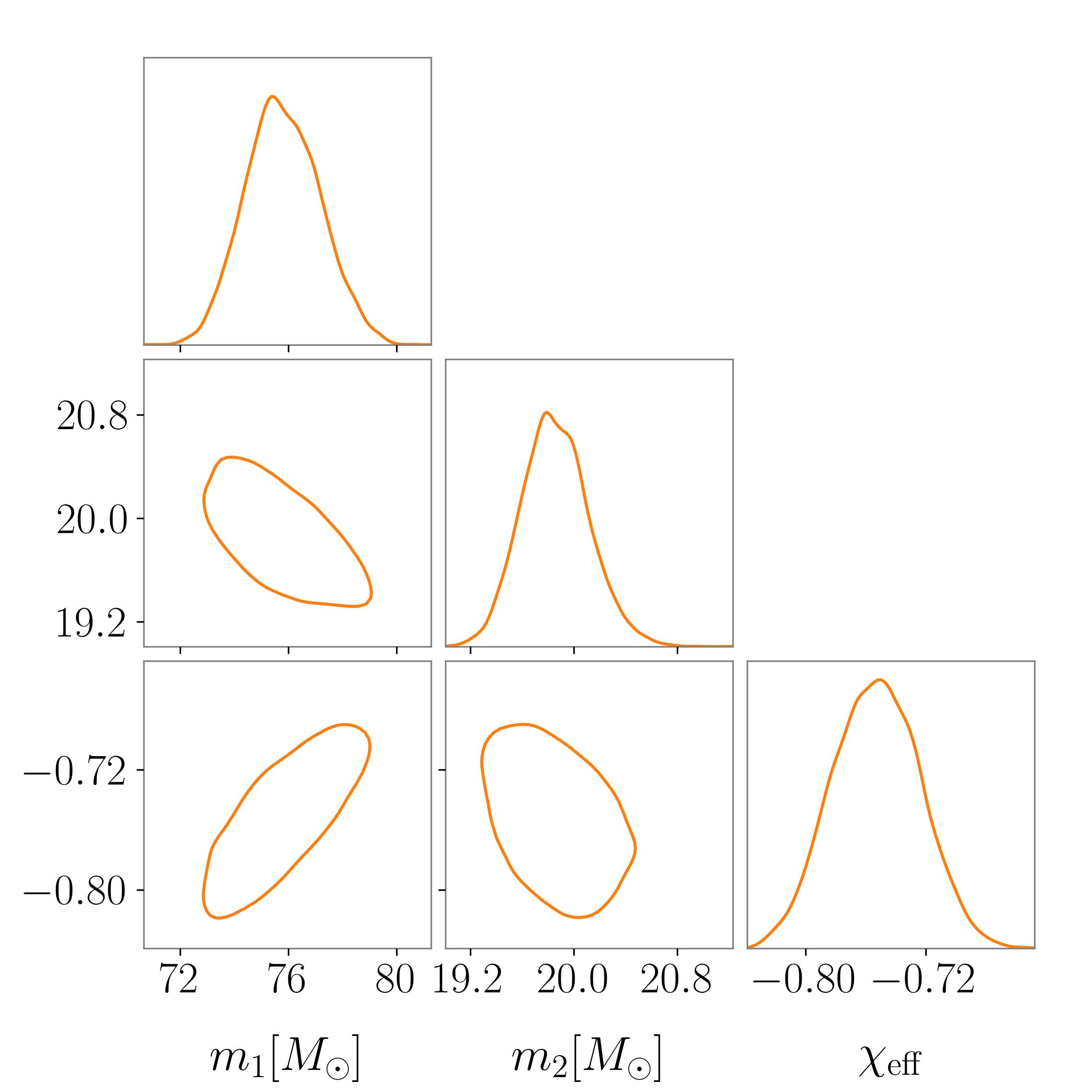}
    \includegraphics[width=0.95\textwidth, trim={0cm 0cm 0cm 1.5cm}, clip]{verification_waveform_posterior.jpeg}
    \caption{The inferred posterior distribution of a simulated signal in non-Gaussian noise. The \emph{Top Left} and \emph{Top right} panels show the inferred component masses and effective aligned-spin obtained under the signal plus noise {\textsc{bilby-antiglitch}} and signal-only {\textsc{bilby}} hypotheses respectively. The vertical and horizontal crosshairs show the true values, and the contours show the $90\%$ credible intervals. Due to significant biases, the \emph{Top Right} panel does not show the true values. The \emph{Bottom} panel shows the the reconstructed GW signals inferred from the {\textsc{bilby-antiglitch}} and {\textsc{bilby}} analyses against the whitened strain data in LIGO-Livingston. The reconstructed GW signals are plotted as a band representing the 90\% credible region.}
    \label{fig:verification_inference}
\end{figure*}

\begin{align} \label{eq:glitch_model}
    \mathcal{G} & = A\exp\left(i\varphi - 2\pi ift_{0}\right) h_{0}(f) \\
    h_{0}(f) &= \mathcal{N}^{-1}\exp\left( -\frac{\gamma}{2}(\log f - \log f_{0})\right) ,
\end{align}
where,

\begin{equation}
    \mathcal{N} = \sqrt{\frac{\langle h_{0}(f) | h_{0}(f) \rangle}{S_{n}(f)}}.
\end{equation}
This phenomenology was tested against glitches found in the third GW observing run. However, {\textsc{AntiGlitch}} also has its disadvantages. Primarily, it has only been designed to handle short duration noise transients, such as tomtes, blips, low-frequency blips and koi fish~\citep{Bondarescu:2023jcx}. This means that it will not model e.g. low-frequency scattering glitches; other models have been designed for this case~\citep{Udall:2022vkv,Tolley:2023umc}. Nevertheless, {\textsc{AntiGlitch}} has been shown to model the majority of glitches identified by GW detectors~\citep{Ashton:2026seh}. 

In this work, we introduce {\textsc{bilby-antiglitch}}. {\textsc{bilby-antiglitch}} is an extension of {\textsc{bilby}} to jointly infer both the signal and noise. As a proof of principle, we implement the {\textsc{AntiGlitch}} model, but like {\textsc{bilby}}, {\textsc{bilby-antiglitch}} is designed to be modular, allowing for alternative glitch models to be added when available. The user is free to choose the optimal glitch model \emph{a priori}, which is possible, since glitch classification is performed ahead of time.

\section{Verification and implications} \label{sec:verification}

First, we verify {\textsc{bilby-antiglitch}} for a synthetic GW signal in the presence of loud glitches. We also demonstrate some of the implications of using a signal-only hypothesis in non-Gaussian data.

For simplicity, we inject a GW150914-like signal simulated with {\texttt{IMRPhenomXPHM}} into idealised Gaussian noise. We consider a binary with masses $m_{1} = 30\, M_{\odot}, m_{2} = 28\, M_{\odot}$ and spins aligned with the orbital angular momentum $\chi_{1} = 0.2, \chi_{2} = 0.2$. All other spin components are exactly zero, implying that there is no spin-precession in the binary. The luminosity distance of the source is chosen so that the signal-to-noise ratio (SNR) is $\rho = 30$. In this analysis, we only consider LIGO-Livingston operating at its design sensitivity since the probability of coincident glitches in a network of interferometers is minimal. To simulate a short duration non-Gaussian artifact, we inject a blip glitch $0.2$ seconds before merger with SNR $\rho=70$. We model the blip glitch using Eq.~\ref{eq:glitch_model} with a central frequency $f_{\mathrm{peak}} = 100\,\mathrm{Hz}$ and an inverse bandwidth squared $\gamma = 5\, \mathrm{Hz}$. Since we are using Eq.~\ref{eq:glitch_model}, we assume a blip phenomenology from the third GW observing run. An omegascan of the simulated signal can be seen in Fig.~\ref{fig:verification_spectrogram}. Given that the glitch is loud and overlaps the simulated signal in both time and frequency, we expect this to be a worst-case scenario with significant biases in our parameter estimation when analysing the data with a signal-only model~\citep{Hourihane:2022doe}. For comparison in the first half of the fourth GW observing run, the majority of glitches had SNR $\rho < 10$~\citep{LIGO:2024kkz}.

\begin{figure}
    \includegraphics[width=0.49\textwidth]{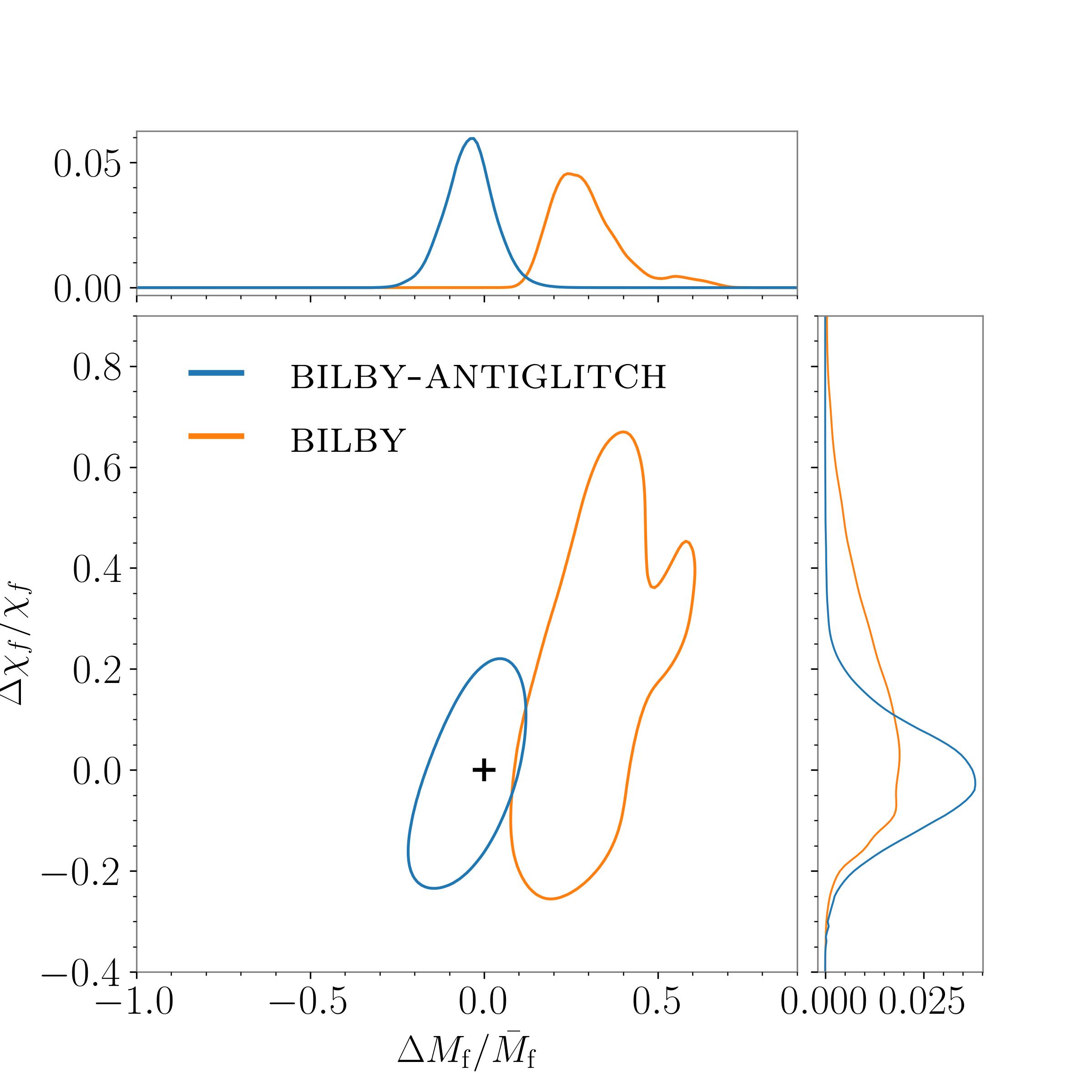}
    \caption{The consistency of a simulated signal in non-Gaussian noise with a binary black hole in General Relativity. We use data from the inspiral and merger-ringdown to construct a posterior distribution for $\Delta M_{f} / \bar{M}_f$ and $\Delta \chi_{f} / \bar{\chi}_{f}$. We compare the results under the signal plus noise hypothesis from {\textsc{bilby-antiglitch}} to the results from the signal-only hypothesis from {\textsc{bilby}}. The contours show the 90\% credible interval and $(0, 0)$ is the expected value for General Relativity.}
    \label{fig:verification_tgr}
\end{figure}

\begin{figure}
    \includegraphics[width=0.48\textwidth]{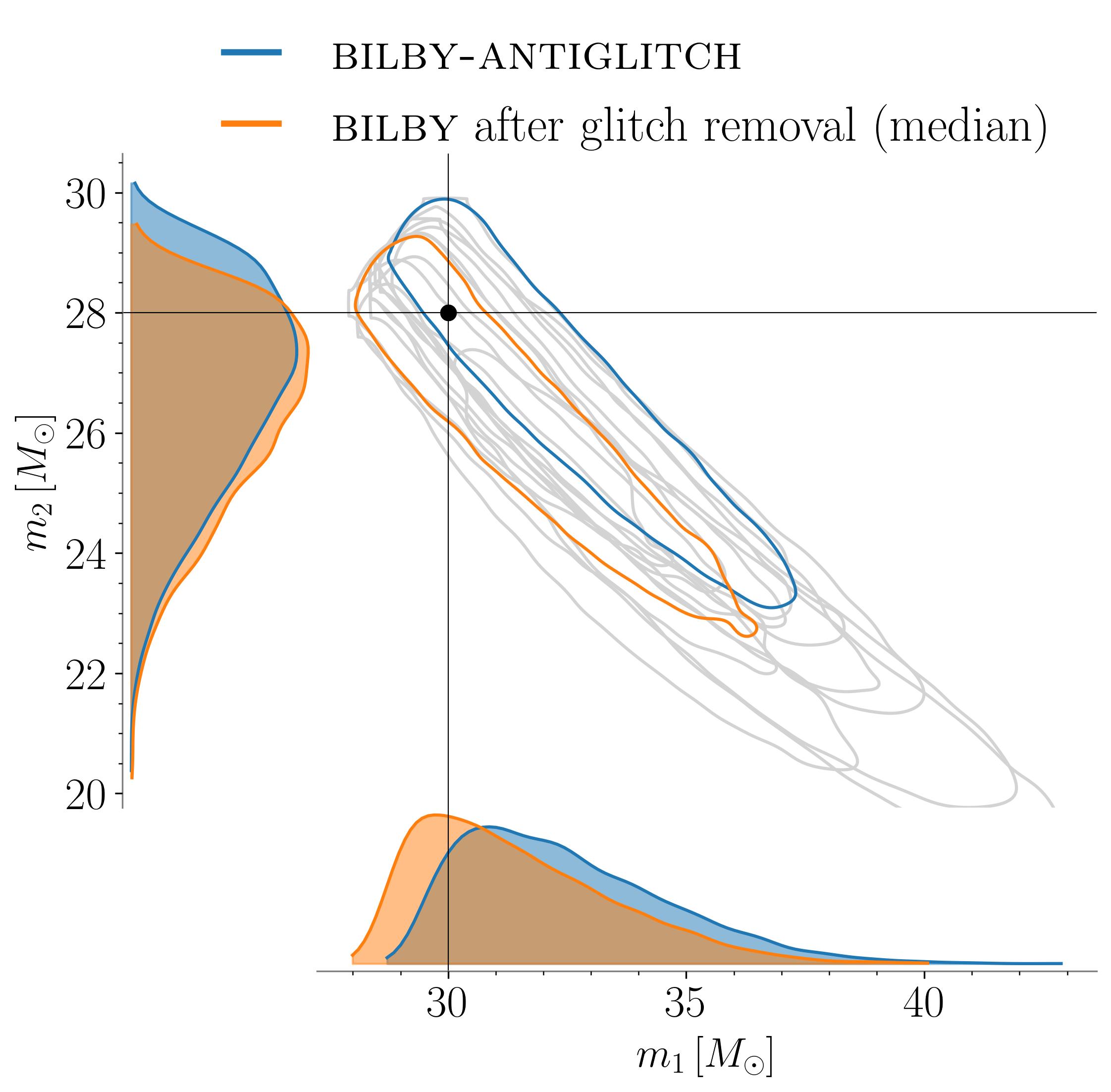}
    \caption{The inferred component masses when analysing a simulated signal in non-Gaussian noise. We compare the inferred distributions from our joint signal plus noise analysis with {\textsc{bilby-antiglitch}}, and from a signal-only analysis with {\textsc{bilby}} after the median realisation of the glitch inferred with {\textsc{bilby-antiglitch}} is subtracted. We also show the variation in the inferred component masses when performing a signal-only analysis after subtracting off random realisations of the glitch in grey.}
    \label{fig:verification_mass_comparison}
\end{figure}

Throughout this analysis, we use wide and agnostic priors for both the glitch and astrophysical signal. For the astrophysical signal we follow the conventions adopted by the LVK. Specifically, we sample uniformly in component masses between a chirp mass $15\,M_{\odot} < \mathcal{M} < 35\, M_{\odot}$ and mass ratio $0.05 < m_{2} / m_{1} < 1$. The luminosity distance is also assumed to be uniform in volume between $10\,\mathrm{Mpc} < d_{\mathrm{L}} < 10000\,\mathrm{Mpc}$. For the glitch, we use uniform priors for the amplitude between $0 < A < 500$, peak frequency between $15 < f_{0} < 1000\,\mathrm{Hz}$ and inverse bandwidth squared $0 < \gamma < 1000\,\mathrm{Hz}$. All other parameters are assumed to be agnostic within their regions of validity. For the astrophysical signal, we recovered with the {\texttt{IMRPhenomXPHM}} model, and for the glitch we used {\textsc{AntiGlitch}}.

In Fig.~\ref{fig:verification_inference} we show the inferred posterior distribution when analysing the data with {\textsc{bilby-antiglitch}}. For comparison, we also show the posterior inferred from the traditional signal-only model with {\textsc{bilby}}. As expected, the non-Gaussian noise transient causes significant biases in the inferred component masses and spin magnitudes when using the signal-only hypothesis; unlike the true source properties, we infer a heavy, asymmetric component mass binary with near-extremal spins that are anti-aligned with the angular momentum of the binary. This is consistent with the signal-only model fitting the noise transient rather than the underlying astrophysical signal, see e.g.~\cite{Ashton:2021tvz}. Indeed, this can be seen in the reconstructed GW signals. However, when the data is analysed under the joint signal plus noise hypothesis with {\textsc{bilby-antiglitch}}, we recover the true source properties of the binary. This is expected since the underlying noise distribution is Gaussian when subtracting both components of the injection. From the reconstructed GW signals, we see that both the glitch and astrophysical signal are recovered with high fidelity. 

When computing the inferred SNRs for both analyses, we see that {\textsc{bilby-antiglitch}} obtains an SNR of $30.8^{+0.1}_{-0.1}$ for the astrophysical signal, and $69.86^{+0.01}_{-0.05}$ for the glitch. This is comparable to the injected values. On the other hand, {\textsc{bilby}} obtains an SNR of $68.9^{+0.1}_{-0.1}$ for the astrophysical signal, comparable with the glitch. When computing the log Bayes factor for signal versus noise, we see that {\textsc{bilby-antiglitch}} out performs {\textsc{bilby}} by a factor of $\sim 10^{250}$; {\textsc{bilby-antiglitch}} obtains a log Bayes factor of $\log_{10}\mathcal{B}_{S/N} = 1244$ compared to $\log_{10}\mathcal{B}_{S/N} = 997$.

In terms of computational cost, we found that {\textsc{bilby-antiglitch}} used $2\times$ less computational cost than {\textsc{bilby}}; {\textsc{bilby-antiglitch}} used 500 CPU hours compared to 940 CPU hours for {\textsc{bilby}}.  This is despite {\textsc{bilby-antiglitch}} sampling over a larger parameter space. This is consistent with {\textsc{bilby}} sampling from a non-trivial likelihood surface, where an astrophysical signal model is attempting to describe a glitch.

\begin{figure*}
    \includegraphics[width=0.95\textwidth]{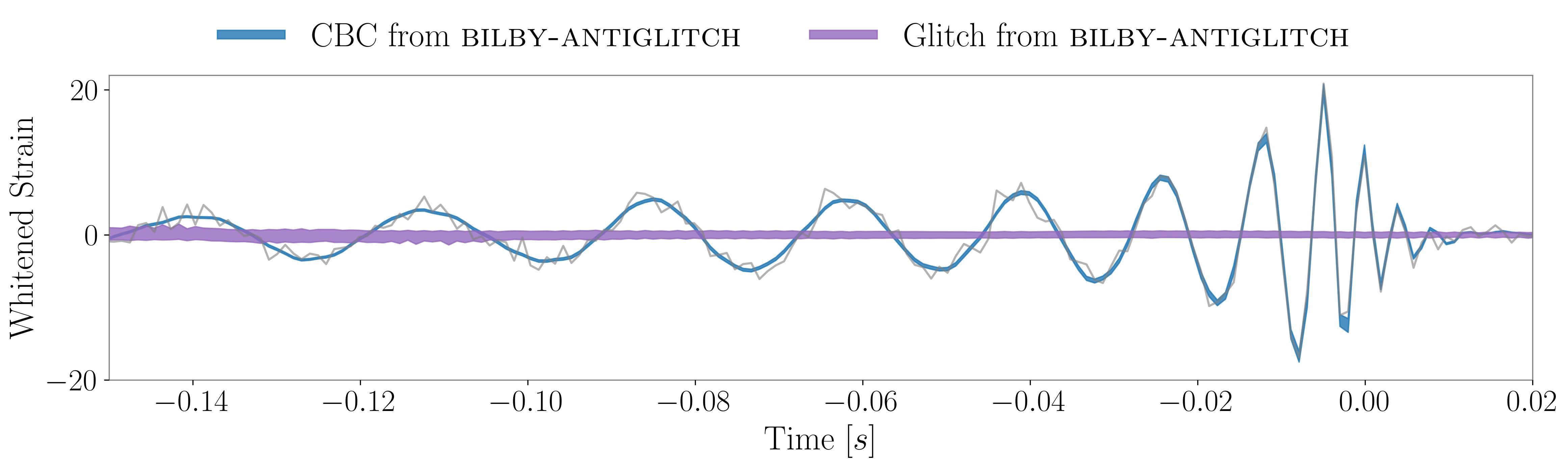}
    \caption{The reconstructed astrophysical and glitch signals inferred by {\textsc{bilby-antiglitch}} when analysing GW250114\_082203. For simplicity we only show the glitch and astrophysical CBC signal projected into LIGO-Livingston. For comparison we show the whitened strain data. The reconstructed GW signals are plotted as a band representing the 90\% credible interval.}
    \label{fig:GW250114_whitened_strain}
\end{figure*}

Given that the signal-only analysis is fitting the glitch rather than the underlying GW signal, we expect significant biases when testing for binary black hole consistency across the waveform; this test has become a standard check for the LVK collaboration~\citep{LIGOScientific:2026qni}. For example, from the amplitude, phase and frequency evolution of the inspiral, estimates for the remnants mass and spin can be obtained through numerical-relativity fits~\citep{Jimenez-Forteza:2016oae, Healy:2016lce, Hofmann:2016yih}. On the other hand, from the frequency and decay time of the ringdown portion of the signal, estimates for the remnants mass and spin can be estimated by assuming the Kerr metric~\citep{Echeverria:1989hg, Finn:1992wt}. Assuming General Relativity is correct, these two independent methods for obtaining the final mass and spin of the remnant should agree~\citep{LIGOScientific:2016lio}. The following dimensionless quantities are used to quantify this agreement~\citep{Ghosh:2016qgn,Ghosh:2017gfp},

\begin{equation}
    \Delta M_{f} / \bar{M}_f,\, \Delta \chi_{f} / \bar{\chi}_{f},
\end{equation}
where an overbar denotes the average between the inspiral (I) and the merger-ringdown (MR) values,

\begin{align}
\bar{M}_f & = \left(M_f^\mathrm{I} + M_f^\mathrm{MR}\right)/{2}, \label{eq:M_f_bar}\\
\bar{\chi}_f &:= \left(\chi_f^{\mathrm{I}} + \chi_f^\mathrm{MR}\right)/{2} \label{eq:a_f_bar}
\end{align}
The transition frequency between the inspiral and the merger and ringdown portion of the signal, is typically chosen to be the Innermost Stable Circular Orbit frequency of an equatorial, prograde timelike orbit around a Kerr black hole~\citep{Ghosh:2016qgn}.

In Fig.~\ref{fig:verification_tgr} we see that the signal-only hypothesis obtained with {\textsc{bilby}} is not consistent with a binary black hole since $\Delta M_{f} / \bar{M}_f$ and $\Delta \chi_{f} / \bar{\chi}_{f}$ excludes $(0, 0)$ at more than 90\% confidence. If this event was real, and in the rare case that the the glitch was not identified \emph{a priori}, this signal would show a deviation from General Relativity. Interestingly, we see that $\Delta \chi_{f} / \bar{\chi}_{f}$ is consistent with $0$, and the remnant mass causes the deviation. As expected, the signal reconstructed with {\text{bilby-antiglitch}} is consistent with a binary black hole, recovering the expectation from General Relativity within the 90\% credible interval.

In reality, the LVK does not perform a signal-only analysis on the data shown in Fig.~\ref{fig:verification_spectrogram}. Rather, glitch mitigation techniques are applied before Bayesian inference is performed~\citep[see e.g.][]{LIGOScientific:2026wfs}. Typically, a preliminary analysis is performed on the glitch with {\textsc{BayesWave}}~\citep{Cornish:2014kda,Littenberg:2015kpb,Cornish:2020dwh,Gupta:2024tve} and the median glitch realisation is subtracted from the data, leaving behind the underlying astrophysical signal for analysis. While {\textsc{bilby-antiglitch}} is designed to perform joint inference, we can reproduce this process, and subtract only the non-Gaussian noise artifact for subsequent analyses to be undertaken with {\textsc{bilby}}. 

In Fig.~\ref{fig:verification_mass_comparison} we show the inferred component masses when analysing our simulated signal with {\textsc{bilby}}, after subtracting the median glitch realisation inferred by {\textsc{bilby-antiglitch}}. For comparison, we also show the inferred posterior when subtracting different realisations of the glitch (rather than the median), and when jointly modeling the signal plus noise. In Appendix~\ref{sec:glitch_subtracted_data}, we also show the spectrogram after subtracting the median glitch and signal realisations. We see that all methods recover the true parameters of the binary. However, we see a large variation in the inferred binary properties when subtracting different glitch realisations, beyond those expected from the stochastic nature of sampling~\citep[see e.g.][]{Romero-Shaw:2020owr}. This highlights one of the main disadvantages with removing a single glitch realisation. We also see small differences between the joint and signal-only inferences, implying that some of the astrophysical signal is being removed by the glitch subtraction. This is the reason why, although possible, we recommend performing a joint signal plus noise inference where possible to marginalize over different glitch realisations.

\section{Application} \label{sec:application}

After verifying that {\textsc{bilby-antiglitch}} can be used to analyse simulated signals in non-Gaussian noise, as well as demonstrating that it can prevent biased posterior distributions when analysing (potentially) realistic data with traditional {\textsc{bilby}}, we now analyse several real GW signals. Throughout this section, we use the same sampler settings, PSDs, calibration envelopes etc. as the original LVK analyses. We also use the {\textsc{AntiGlitch}} model to describe any non-Gaussian noise transients. While this may not be the optimal model for the real GW data considered in this section, we highlight that this is a proof of principle, and additional glitch models will be implemented in the future.

\subsection{GW250114\_082203}

GW250114\_082203~\citep{LIGOScientific:2025rid} was observed in the second half of the fourth GW observing run~\citep{LIGOScientific:2026wfs}. It is the loudest GW signal detected to date, and was inferred to have source properties similar to GW150914~\citep{LIGOScientific:2016aoc,LIGOScientific:2016vlm}. At the time of this event, there were no data quality issues reported in LIGO-Hanford, but there was some excess power in LIGO-Livingston at low frequencies. This is therefore an ideal candidate to test {\textsc{bilby-antiglitch}} in the high SNR regime. In our re-analysis, we use the NRSur7dq4 waveform model~\citep{Varma:2019csw} due to its improved accuracy.

In general, we obtain consistent results with the LVK: for the astrophysical signal, we infer an equal mass ratio binary with $m_{2} / m_{1} = 0.95^{+0.05}_{-0.09}$ and a chirp mass $\mathcal{M} = 31.2^{+0.4}_{-0.6}\, M_{\odot}$. In Fig.~\ref{fig:GW250114_whitened_strain}, we show the reconstructed GW signals obtained with {\textsc{bilby-antiglitch}}. We see remarkable agreement between the CBC signal and the underlying whitened data. Interestingly, we see small, but non-zero, vibrations inferred by the glitch model in LIGO-Livingston. This correlates with the model trying to capture the reported excess low-frequency power. As expected, we obtain no evidence for non-Gaussian noise artifacts in LIGO-Hanford. 

Regarding computational cost, we found that {\textsc{bilby-antiglitch}} used $1.5\times$ more computational resources than {\textsc{bilby}}. This is expected given the larger parameter space. Unlike in Section~\ref{sec:verification}, this is because there is no strong evidence for a glitch in the data, meaning that the likelihood surface is not made more difficult to sample when using the signal-only hypothesis.

From this analysis we see that {\textsc{bilby-antiglitch}} is robust to analysing real GW signals, even when there are no significant non-Gaussian noise transients reported. Although further testing is needed, this implies that it can generally be used for the majority of GW signals reported by the LVK. The only downside is an increase in the computational cost due to the larger parameter space explored.

\subsection{GW200129\_065458}

Next, we consider GW200129\_065458~\citep{KAGRA:2021vkt}. GW200129\_065458 was observed in the second half of the third GW observing run, and this signal sparked interest within the community as it was the first signal to show significant evidence for spin-precession~\citep{Hannam:2021pit}; initial analyses conducted by the LVK obtained different conclusions regarding the evidence for spin-precession in the signal, with one model showing evidence for spin-precession, and another showing no evidence~\citep{KAGRA:2021vkt}. However, the exceptional nature of GW200129\_065458 was later questioned, since it coincided with excess noise caused by the $45\, \mathrm{MHz}$ electro-optic modulator system~\citep{LIGOScientific:2016gtq}. Therefore, the question was asked whether non-Gaussian noise artifacts were causing biased inferences~\citep{Payne:2022spz,Macas:2023wiw}. Here, we re-analyse GW200129\_065458 with {\textsc{bilby-antiglitch}}. Unlike the original LVK analysis, we employ the NRSur7dq4 waveform model since it was shown to more accurately describe General Relativity in this region of the parameter space~\citep{Hannam:2021pit}.

\begin{figure}
    \hspace{-2em}
    \includegraphics[width=0.55\textwidth]{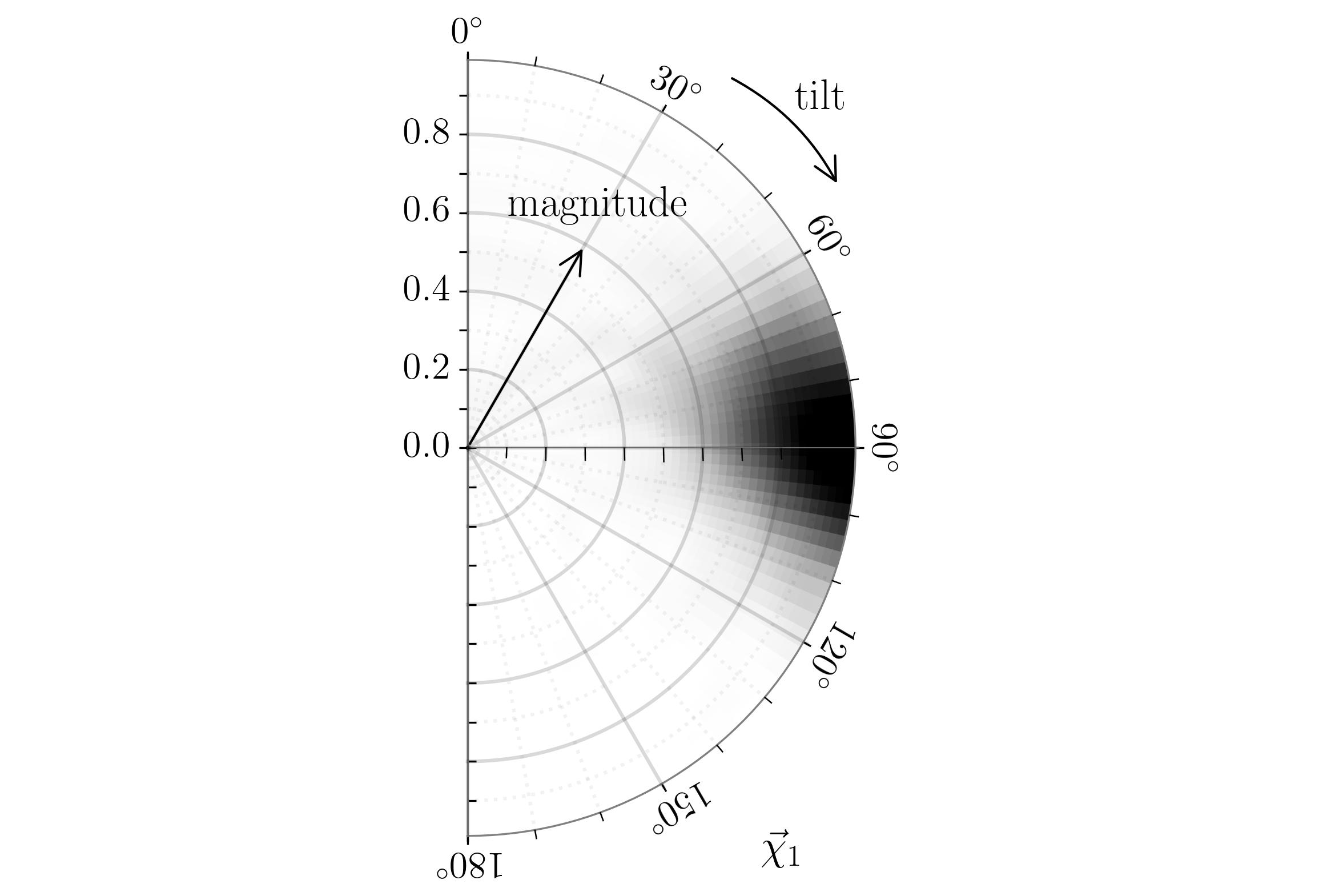}
    \caption{The two-dimensional posterior distribution for the spin of the primary black hole when analysing GW200129\_065458 with {\textsc{bilby-antiglitch}}. Tilt angles of $90^\circ$ means that the spin vector lies within the plane of the binary. The colour indicates the posterior probability per pixel. This plot is produced by using histogram bins that are constructed linearly in spin magnitude and the cosine of the tilt angles such that each bin contains identical prior probability. The probabilities are marginalized over the azimuthal angles.
    }
    \label{fig:GW200129_spin_disk}
\end{figure}

In Fig.~\ref{fig:GW200129_spin_disk}, we show the two dimensional probability distribution for the spin of the larger black hole. We see that the spin lies almost within the plane of the binary, indicating maximal spin-precession. Our results are consistent with those found in~\cite{Hannam:2021pit, Macas:2023wiw,Islam:2023zzj}, and contrast those found in~\cite{Payne:2022spz} where {\textsc{BayesWave}} was used to model the noise. Based on our analysis, we find that the precession measurement of GW200129\_065458 remains robust to short duration non-Gaussian noise transients. 

\section{Conclusion} \label{sec:discussion}

In this work, we combined the core joint signal plus noise inference ideology presented in {\textsc{BayesWave}} with the LVK flagship Bayesian inference pipeline {\textsc{bilby}} to provide an alternative algorithm for mitigating the effect of short duration non-Gaussian noise transients on Bayesian parameter estimation. We demonstrated that, unlike traditional techniques, our algorithm successfully recovered the true source properties of an astrophysical signal in non-Gaussian noise. We further demonstrated that it can help to prevent false violation claims of General Relativity due to non-Gaussianities in the data. Finally, although not recommended, we also demonstrated that {\textsc{bilby-antiglitch}} can be used within the current infrastructure by producing glitch subtracted data for traditional techniques to analyse.

In this work, we only implemented a quasi-physical parameterised glitch model, suitable for short-duration non-Gaussian noise transients, as a proof of principle. Since {\textsc{bilby-antiglitch}} is designed to be modular, additional glitch models can trivially be added~\citep[see e.g.][]{Merritt:2021xwh,Udall:2022vkv,Tolley:2023umc}. Once further glitch models are included, a more in depth investigation detailing the effect of non-Gaussian noise on Bayesian analyses can be conducted.

Since {\textsc{bilby-antiglitch}} builds upon the LVK's flagship Bayesian inference library, it can also be used alongside all of the latest waveform models, likelihood acceleration algorithms~\citep{Cornish:2010kf,Canizares:2014fya,Vinciguerra:2017ngf,Zackay:2018qdy,Morisaki:2020oqk,Cornish:2021lje,Morisaki:2021ngj,Krishna:2023bug,Prathaban:2026kft}, waveform accuracy techniques~\citep{Hoy:2024vpc} and samplers. Similarly, although the expected glitches in next generation space-based detectors~\citep{LISA:2017pwj} will likely be much longer in duration, our framework can be easily incorporated into the {\textsc{bilby-lisa}} package~\citep{Hoy:2023ndx}, and additional, next-generation specific glitch models, can be incorporated.

\section{Acknowledgments}

We thank Ann-Kristin Malz for comments during the LIGO--Virgo--KAGRA internal review, as well as Christopher Berry, Shun Yin Cheung, Eric Thrane and Rhiannon Udall for discussions. C.H thanks the University of Portsmouth for support through the Dennis Sciama Fellowship. R.B acknowledges financial support from the grants RYC2022-035983-I, PID2024-159689NB-C22, and CEX2024-001451-M funded by MCIN/AEI/10.13039/501100011033 (State Agency for research of the Spanish Ministry of Science and Innovation) and SGR-2021-01069 (AGAUR). L.K.N thanks the UKRI Future Leaders Fellowship for support through the grant UKRI2745. We are also grateful for computational resources provided by the SCIAMA high perfomance computing cluster which is supported by the Institute of Cosmology and Gravitation (ICG) and the University of Portsmouth.

This research has made use of data or software obtained from the Gravitational Wave Open Science Center (gwosc.org), a service of the LIGO Scientific Collaboration, the Virgo Collaboration, and KAGRA. This material is based upon work supported by NSF's LIGO Laboratory which is a major facility fully funded by the National Science Foundation, as well as the Science and Technology Facilities Council (STFC) of the United Kingdom, the Max-Planck-Society (MPS), and the State of Niedersachsen/Germany for support of the construction of Advanced LIGO and construction and operation of the GEO600 detector. Additional support for Advanced LIGO was provided by the Australian Research Council. Virgo is funded, through the European Gravitational Observatory (EGO), by the French Centre National de Recherche Scientifique (CNRS), the Italian Istituto Nazionale di Fisica Nucleare (INFN) and the Dutch Nikhef, with contributions by institutions from Belgium, Germany, Greece, Hungary, Ireland, Japan, Monaco, Poland, Portugal, Spain. KAGRA is supported by Ministry of Education, Culture, Sports, Science and Technology (MEXT), Japan Society for the Promotion of Science (JSPS) in Japan; National Research Foundation (NRF) and Ministry of Science and ICT (MSIT) in Korea; Academia Sinica (AS) and National Science and Technology Council (NSTC) in Taiwan. For the purpose of open access, the author(s) has applied a Creative Commons Attribution (CC BY) licence to any Author Accepted Manuscript version arising.

\section{Data Availability}

The {\textsc{bilby-antiglitch}} software is publicly available on \href{https://github.com/hoyc1/bilby_antiglitch}{GitHub} to reproduce this work.

\appendix

\section{Glitch subtracted data} \label{sec:glitch_subtracted_data}

\begin{figure}
    \includegraphics[width=0.49\textwidth]{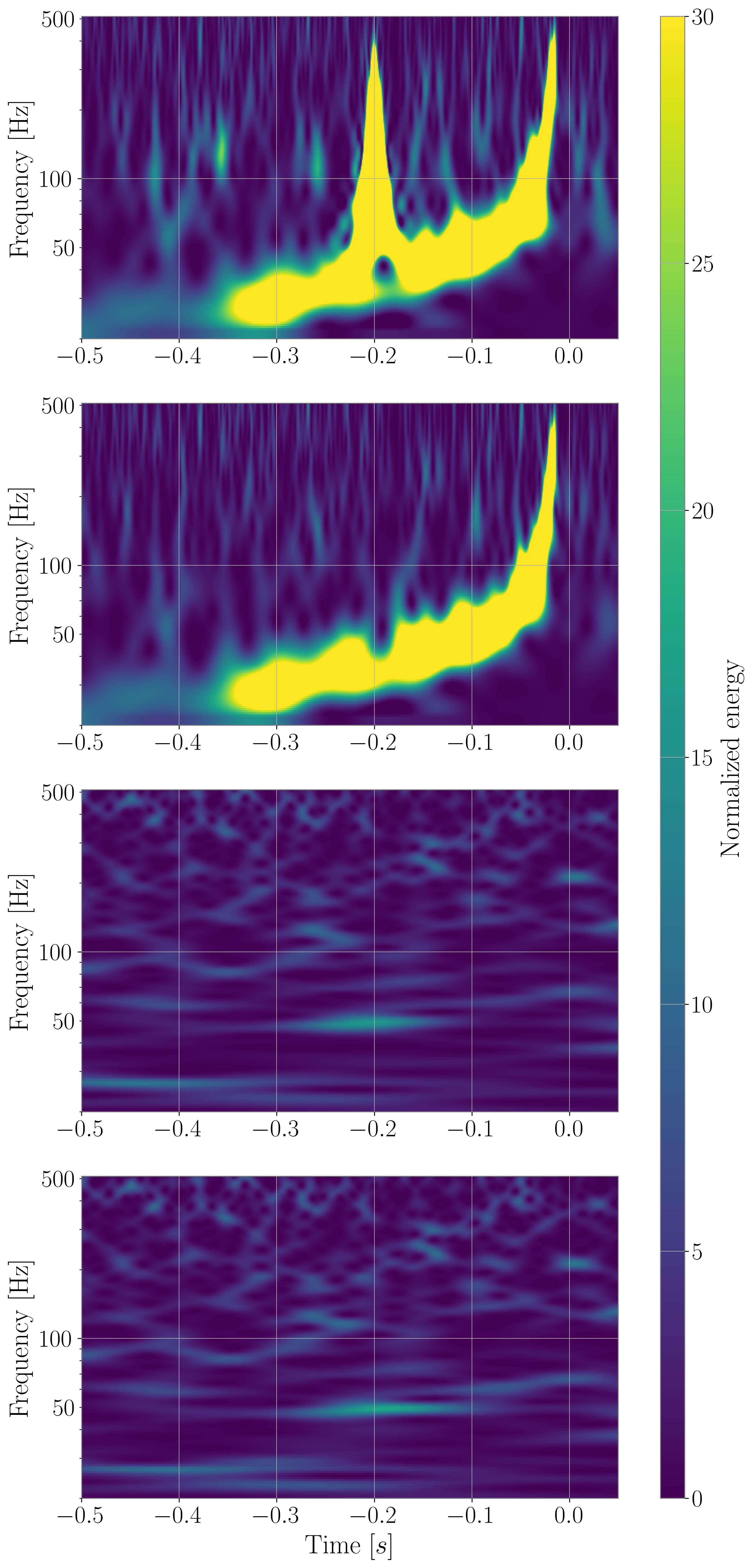}
    \caption{The spectrogram of a simulated GW signal in the presence of non-Gaussian noise, after glitch and signal subtraction is applied. The \emph{First} panel is a reproduction of Fig.~\ref{fig:verification_spectrogram}. The \emph{Second} panel shows the spectrogram after subtracting the median glitch realisation inferred by {\textsc{bilby-antiglitch}}. This corresponds to the data analysed by {\textsc{bilby}} in Fig.~\ref{fig:verification_mass_comparison}. The \emph{Third} panel shows the spectrogram after additionally subtracting the median signal realisation inferred by {\textsc{bilby}}, when analysing the data in the \emph{Second} panel. Finally, the \emph{Fourth} panel shows the spectrogram after subtracting the median signal plus glitch realisation inferred by {\textsc{bilby-antiglitch}}, when analysing the data in the \emph{First} panel.}
    \label{fig:verification_spectrogram_subtracted_frames}
\end{figure}

In Section~\ref{sec:verification}, we highlighted that in reality, the LVK does not perform a signal-only analysis on data containing non-Gaussian noise transients. Instead, a signal-only analysis is performed on data where the median glitch realisation is subtracted. In Fig.~\ref{fig:verification_spectrogram_subtracted_frames}, we show the spectrograms of the data before, and after the median glitch and signal realisations have been subtracted. We see that when {\textsc{bilby-antiglitch}} subtracts the median glitch and signal joint realisation inferred from the original non-Gaussian data, the resulting background is Gaussian as expected. This implies that the fundamental assumption of the Whittle likelihood is valid. Similarly, we see that when {\textsc{bilby-antiglitch}} subtracts the median glitch realisation inferred from the original non-Gaussian data, we obtain a clean astrophysical signal for {\textsc{bilby}} to analyse. As expected, we see that when {\textsc{bilby}} subtracts the median signal realisation inferred from the glitch subtracted data, we obtain the expected Gaussian assumption.

\bibliographystyle{mnras}
\bibliography{main}

\label{lastpage}
\end{document}